\documentclass[aps,prd,preprint]{revtex4}

\newcommand{\beq}{\begin{equation}}
\newcommand{\eeq}{\end{equation}}
\newcommand{\beqa}{\begin{eqnarray}}
\newcommand{\eeqa}{\end{eqnarray}}
\newcommand{\beqar}{\begin{eqnarray*}}
\newcommand{\eeqar}{\end{eqnarray*}}

\newtheorem{theorem}{Theorem}
\newtheorem{corollary}{Corollary}

\def \B {\beta_o e^{\Phi}}
\def \Bo {\beta_o}

\def \s {\,\,\,\,}

\def \e {{\bf e}}
\def \t {\hat{t}}
\def \r {\hat{r}}

\begin{document}


\title{A Thermodynamic Sector of Quantum Gravity}


\author{Jonathan Oppenheim}
\affiliation{Theoretical Physics Institute, University of Alberta, 412 Avadh Bhatia
Physics Laboratory, Alta., Canada, T6G 2J1.}
\email[]{jono@phys.ualberta.ca}

\date{\today}

\begin{abstract}
The connection between gravity and thermodynamics is explored.
Examining a perfect fluid
in gravitational equilibrium we find that the entropy is extremal only
if Einstein's equations are satisfied.  Conversely, one can derive part
of Einstein's equations from
ordinary thermodynamical considerations.  This allows the theory of
this system to be recast in such a way that a sector of general relativity is
purely thermodynamical and should not be quantized.
\end{abstract}

\pacs{04.70.Dy, 04.40.-b, 05.70.-a}

\maketitle
\section{introduction}

Einstein's equations relate the geometry of space-time (as expressed
by the Einstein tensor $G_{\mu\nu}$) to the distribution of matter
(as expressed by the stress-energy tensor $T_{\mu\nu}$ ).
%
%
Many connections between these equations and thermodynamics have been
emerging in recent years.  Einstein's equations imply that black holes possess a
temperature\cite{hawking}, and their horizon behaves like an entropy
\cite{jacob}. In dynamical
gravitating systems, critical phenomena has been observed
\cite{matt}.

In these examples, the horizon seems to play a central role.
In this paper, we will explore the connection between thermodynamics
and gravity by examining a gravitating system which does not possess a
horizon.  In particular, a spherically
symmetric perfect fluid in equilibrium will be considered.
We find that the connection between the thermodynamics of this system
and general relativity is extremely deep - in fact, it is an identity.
By extremizing the total entropy
of the matter while keeping the total particle number and energy fixed, one
finds that the most probably configuration is one which obeys a linear
combination of the two independent Einstein's equations (the other 8 components
of Einstein's equations are trivially satisfied).
The fact that the entropy of the system is only an extremum when Einstein's
equations are satisfied can be considered as a counterpart to black hole
thermodynamics for systems without horizons.

%

However, this close connection between thermodynamics and
Einstein's equations  appear somewhat
puzzling  -- there does not appear to be any reason {\it a priori} why
the entropy should only be an extremum for Einstein's equations.
This leads us to suggest that this can be viewed as evidence
that Einstein's equations represent an effective theory.
In particular, for this system, one can
derive one of Einstein's equations from thermodynamics.

As a result
one need not
assume all of Einstein's equations for this system.  Rather, one only needs to
assume a single equation, and the remaining equations will be automatically
satisfied due to thermodynamical considerations. One can therefore view
one of the two independent Einstein's equations as having no
physical content - it is just a restatement of the laws of equilibrium
statistical mechanics in curved space.

If one adopts such a position, then one need not quantize the equation
which is derived from thermodynamics.  In fact, it would make as much
sense to quantize
this equation as it would be to quantize the laws of hydrodynamics or
the laws of thermodynamics themselves.  One does not quantize the Navier-Stokes
equation or the first law of thermodynamics; rather, one quantizes the
underlying equations governing the individual molecules which make up the fluid
or thermodynamic system.

The notion that general relativity is an effective theory which should
not be quantized directly was strongly supported by the derivation
given by Jacobson \cite{ted}.  There, it was shown that if there exists a
microscopic theory where causal horizons have an entropy proportional to their
area, then Einstein's equations will automatically hold.  In the present work,
no assumption need be made about the underlying microscopic theory of geometry.
In fact, here, the quantum theory is simply the ordinary quantum theory of the
matter.

The thermodynamics and statistical mechanics of gravitating systems
have been studied using toy models, mean field theories,
and numerical methods\cite{paddy}.
In some of these toy models, the entropy is not a global extremum\cite{antanov}
while in other models, and using other methods it is\cite{lynden2}.
For example, if one couples a gravitational system to an infinite and external
thermal reservoir, then the negative specific heat of the gravitational system
will make thermal equilibrium impossible.
Here, there is no external heat bath or containment box, and furthermore, our
study is fully relativistic.  We find that the total entropy is an extremum when
Einstein's equation is satisfied. In general, the study of thermodynamics for
systems with long range interactions can be problematic.  In fact, an entirely
new definition of entropy has been used to attempt to deal with these
systems\cite{tsallis} because it was not believed that one can properly analyze
these systems using conventional thermodynamics. Here, we see that using the
equivalence principle to define all thermodynamic quantities locally allows us
to completely analyze the system without any need to make approximations or go
outside traditional thermodynamics. A more exhaustive analysis of these issues
will be given elsewhere \cite{inprep}.

In the next section, we will introduce the system under consideration
and show that local validity of the first law of thermodynamics
is equivalent to the vanishing of the divergence of the stress-energy tensor.
This is not at all surprising but the
method of derivation is instructive.
We will then derive the conditions for maximal entropy and show
that it is equivalent to one of
Einstein's equations.
In Section \ref{therm} we show that for this
system one can recast the equations of general relativity into a
different form where one equation is purely thermodynamical and the other
equation gives the physical theory.  We
conclude with a discussion on the possible implications of this result for the
quantum theory of gravity.

Note: I have recently become aware of
a related work by Harrison et. al. \cite{red} in their study of
gravitating fluids. There, the entropy of the system is assumed to be
zero, however the methodology is somewhat
similar.  They assume the correct form for the proper volume in the space
time of a perfect fluid.  They then derive the Oppenheimer-Volkoff equation
by minimizing the Arnowitt-Desser-Misner mass. There too, the
authors appear puzzled by this result, although they speculate that that it may
be due to the simplified geometry.

\section{Entropy and Einstein's equations}
Initially, we will not assume any of Einstein's equations.
However, we assume a metric theory of gravity, which will allow us to exploit
the equivalence principle.
Since we are considering a spherically symmetric
system, the metric takes the familiar form
\beq
ds^2=-e^{2\Phi}dt^2 + e^{2\Lambda} dr^2 + r^2 d\Omega^2 \s.
\eeq
Since we are considering a system in equilibrium the metric is independent
of time.  We will not specify the metric functions -- in particular,
we will not demand that this metric satisfy Einstein's equations.
For matter, we assume the
stress-energy tensor of the perfect fluid, which is given in terms of the energy
density $\rho(r)$ and radial pressure $p(r)$ by \beq
T^{\mu\nu}=(\rho+p)u^\mu(r) u^\nu(r) + p(r) g^{\mu\nu}
\eeq
where $u^\mu(r)$ is the 4-velocity of the fluid and $g^{\mu\nu}$ is the metric.
It is worth noting that spherical symmetry would follow automatically for a
perfect fluid in equilibrium (i.e. stationary) if we were to assume asymptotic
flatness.

We make no assumptions about specific equations of states for
the matter, but we will assume that in the absence of gravity, the fluid
has no unscreened long-range interactions and is therefore {\it extensive}. In
other words, the entropy, particle number and energy scale as the size of the
system when the intensive variables (temperature, pressure and chemical
potential) are held fixed. In virtually all applications of thermodynamics this
is assumed to be the case (at least implicitly), for without it, taking the
thermodynamic limit becomes problematic and also, no canonical ensemble
would exist\cite{paddy}.  This assumption of extensivity is only assumed to
hold true in the absence of gravity; once we include gravity, it will no
longer hold globally.  In the absence of gravity, extensivity plus the first
law imply the Gibbs-Duhem relation \cite{gd}
\beq
\rho=Ts-p+\mu n
\label{eq:gd}
\eeq
where $s$,$\rho$,$n$ are the entropy density, energy density
and particle number density of the matter as measured in its proper
reference frame, and $T$, $\mu$ and $p$ are the local temperature, chemical
potential and pressure.

A gravitating perfect fluid obeys the three equations\cite{ov}
\beq
\frac{dp}{dr}=-(\rho+p)\frac{d\Phi}{dr}
\label{eq:bianchi}
\eeq
\beq
e^{-2\Lambda}=1-2m(r)/r \label{eq:gtt}
\eeq
\beq
\frac{d\Phi}{dr}=
\frac{m+4\pi r^3 p}{r(r-2m)}\s .\label{eq:grr}
\eeq
%
The first equation follows from the vanishing of
$T^{\mu\nu}_{\,\,\,\,\,;\nu}$, while the second and third equation
are Einstein's equation written in the orthonormal reference frame with tetrads
given by,
\beq
\e_{\t}=\frac{\partial}{e^\Phi\partial t},\,
\e_{\r}=\frac{\partial}{e^\Lambda\partial r},\,
\e_{\hat{\theta}}=\frac{\partial}{r\partial
\theta},\,
\e_{\hat{\phi}}=\frac{\partial}
{r\sin{\theta}\partial \phi}\s,
\eeq
Equation (\ref{eq:gtt}) is equivalent to the
constraint on initial data
\beq
{^{3}R}=16\pi\rho \label{eq:constraint}
\eeq
where ${^{3}R}$ is the scalar curvature of the intrinsic
geometry defined by
$\e_{\hat r}$,
$\e_{\hat\theta}$, and $\e_{\hat\phi}$.

We will now show that two of these
three equations can be derived from thermodynamics.
Normally, the thermodynamical behavior of
self-interacting theories can be problematic, but in the rest frame of the
perfect fluid we can use all the laws of thermodynamics, and then exploit
the principle of equivalence. In
particular, we
can use the Gibbs-Duhem relation, since
this equation involves local quantities which are scalar fields.   Equation
(\ref{eq:gd}) is therefore valid in all reference frames.  It should be
noted that while the system obeys the Gibbs-Duhem locally, it does not
necessarily obey it globally \cite{scaling}.

Since the system is in thermal
and chemical equilibrium, we can use the
Tolman
relation \cite{tolman}, which tells us that the temperature
and chemical potential at any two points are related by the redshift
factor.  In terms of quantities at infinity $T_o$, $\mu_o$, this gives
us
\beq
T(r)=T_o e^{-\Phi(r)} \label{eq:tolman}
\eeq
and
\beq
\mu(r)=\mu_o e^{-\Phi(r)} \s . \label{eq:chempot}
\eeq
These relations are
purely geometrical and do not depend on Einstein's
equations.  The Tolman relation is simply a reflection
of the fact that frequencies and energies get shifted by a factor
equal to the norm of the static killing vector.

We can now derive equation (\ref{eq:bianchi}) which (not surprisingly)
follow directly from the first law of thermodynamics.  While this result is
trivial, the derivation we will employ is rather interesting in that we will
use the thermodynamical relations  (\ref{eq:gd}),
(\ref{eq:tolman}),(\ref{eq:chempot}), without using any other machinery of
differential geometry.  In other words, local extensivity, plus the Tolman
relation encode all the necessary information.  It is also instructive
as it demonstrates the validity of locally defining the temperature
and other thermodynamic quantities (which can be defined via the first law).

Combining equations (\ref{eq:gd}), (\ref{eq:tolman}) and
(\ref{eq:chempot}) we can write the entropy density as
\beq
s=\beta_o e^\Phi(\rho+p)-\mu_o\beta_o n \label{eq:gdcurved}
\eeq
where $\beta_o$ is just the inverse temperature at
infinity.
Taking the derivative of this equation with respect to $r$
we find
\beq
\frac{ds}{dr}=\beta_o e^\Phi(\frac{d\rho}{dr} +
\frac{dp}{dr})
+
\beta_o \frac{d\Phi}{dr}e^\Phi(\rho+p)
-
\beta_o\mu_o \frac{dn}{dr}
\eeq
and
using the first law
\beq
d\rho=Tds -\mu dn
\eeq
gives us
\beq
\frac{dp}{dr}=-(\rho+p)\frac{d\Phi}{dr}
\label{eq:bianchi2}
\eeq
as required.

We now derive one of Einstein's equations by extremizing the
total entropy of the matter. We will perform the variation at fixed
particle number $N$ and fixed energy at infinity.  For the
latter, we will assume that the energy at infinity is given
by the ADM mass
\beq
M=\int_0^\infty 4\pi r^2 \rho dr \s .\label{eq:adm}
\eeq
This equation clearly has physical content, however, it is
not equivalent to assuming the full Einstein's equation.  It is a
single integral equation and from it we will derive an
equation for the metric which is valid throughout the space
time i.e. from a single equation, we derive an infinite number
of equations, one at every point in space.  In our derivation,
we don't need to assume that the quantity given in equation (\ref{eq:adm})
is actually the total energy, just that it is fixed during the
variation.  However, in order to make contact with
thermodynamics, one would need to identify $M$ with the total energy.

The quantity $N$ is simply given by integrating the local particle
number density $n$ over the proper volume element $dV=4\pi r^2 e^\Lambda dr$

\beq
N=\int_0^\infty n 4\pi r^2 e^\Lambda dr \s ,
\label{eq:number}
\eeq
Unlike equation (\ref{eq:adm}), the equation for $N$ does not rely on
any assumptions other than the equivalence principle.

Before stating the theorems we will prove, we first introduce
the following useful
change of variables
\beq
\frac{dm}{dr}\equiv 4\pi r^2\rho \label{eq:m} \s .
\eeq
This is simply a definition, and no assumptions need be made
concerning the physical meaning of $m$.
In particular, we do not assume that $m$ is given by equation (\ref{eq:gtt}).

We now prove the following:
\begin{theorem}
If we assume that $\Lambda$ is an arbitrary function of $m$ and $r$,
then
\beq
4\pi r^2 (\rho+p)\frac{\partial \Lambda}{\partial m}
=\frac{d\Lambda}{dr}+\frac{d\Phi}{dr}
\eeq
This equation is a linear combination of of Einstein's equations
(\ref{eq:gtt})
-(\ref{eq:grr}) \label{th1}.
\end{theorem}
\begin{corollary}
If Einstein's equations (\ref{eq:gtt}) and (\ref{eq:grr}) are satisfied, then
the entropy is an extremum. \label{cor1}
\end{corollary}
\begin{corollary}
If the equation of constraint (\ref{eq:constraint}) holds and the
entropy is an extremum, then the rest of Einstein's equations hold.
Put another way, if we assume the $tt$ component of Einstein's
equations, which gives us $\Lambda$, then we can derive the
$rr$ component of Einstein's equations which gives us $\Phi$.
\label{cor2}
\end{corollary}

We  now prove Theorem \ref{th1}
by varying the total entropy of the system, to find the extremum.

Since $s$ is the local
entropy as measured by observers in the rest frame of the
fluid, we can integrate this quantity over the proper volume to obtain the
total entropy $S$.
\beq
S
=
\int_0^\infty s 4\pi r^2 e^\Lambda dr
\label{eq:svsr}
\eeq
%
%
We can then append equation (\ref{eq:number}) to the total entropy as a
constraint on the total particle number by
defining a new quantity $L$
\beq
L=S+\lambda\left( \int_0^\infty 4\pi r^2 ne^\Lambda dr
- N\right) \nonumber\\
\eeq
The constraint on the total energy could also be appended
as a constraint, however, this is not necessary as it arises
quite naturally through the variation principle.  Since $\Lambda$ depends on $m$
and $r$, we can simply vary it with respect to $m$, keeping the variation fixed
at the end points.  The vanishing of $\delta\!m$ at the endpoints is equivalent
to varying the entropy at fixed total energy.

We can proceed with the variation by treating
the quantities $\rho(r)$ and $n(r)$ as the independent
thermodynamical variables. We can then extremize the entropy, by varying it with
respect to these quantities in much the same way as a Lagrangian
is varied with respect to conjugate variables $q(t)$ and $\dot{q}(t)$.
We find
\beq \delta \! L =
\int_0^R  4\pi r^2 dr e^\Lambda
\left(\B\delta\!\rho
+(\lambda-\mu_o\Bo) \delta\!n
+(s(\rho,n)+\lambda n)\delta\!\Lambda
\right)
\label{eq:variation}
%
%
\eeq
where we have used the thermodynamical relations
\beq
\beta=\B=\left( \frac{\partial
s}{\partial \rho}\right)_{n}
\eeq
and
\beq
\mu\beta=\mu_o\Bo=-\left(\frac{\partial
s}{\partial n}\right)_{\rho} \s .
\eeq
We can rewrite $\delta\!\rho$ from our definition of $m$ given in
equation (\ref{eq:m})
\beq
\delta \! \rho = \delta \dot{m} /(4\pi r^2)
\eeq
where $\dot{}\equiv\frac{d}{d r}$.
Substituting this and equation (\ref{eq:gdcurved})
into (\ref{eq:variation}) and
then integrating by parts gives
\beqa
\delta\! S&=&\int_0^R dr
\Bo e^{\Lambda+\Phi}
\left(
4\pi r^2(\rho+p)\delta\!\Lambda
- (\dot{\Lambda}+\dot{\Phi})\delta \! m
\right)
\nonumber\\
&+&
\int_0^R 4\pi r^2 dr
e^{\Lambda}
(\lambda-\mu_o\Bo)
( \delta\!n
+
n\delta\!\Lambda)
\label{eq:variation2}
\eeqa
where we have held the variation fixed at the endpoints.  As stated
previously, the vanishing of $\delta\! m(R)=\delta\! M$ is equivalent to
performing the variation at fixed energy, while the vanishing of $\delta\! m(0)$
is a necessary boundary condition to keep $\rho(0)$ finite.
We can then substitute the relation
\beq
\delta \! \Lambda(m,r) = \frac{\partial\Lambda}{\partial m}\delta\! m
\eeq
into Eq. (\ref{eq:variation2}) and since the
system is in equilibrium, the variation of the total
entropy must vanish.  The variation is independent at each point
$r$, and so, the vanishing of $\delta\! L$ implies
\beq
\Bo\mu_o = \lambda
\eeq
and
\beq
4\pi r^2 (\rho+p)\frac{\partial \Lambda}{\partial m}
=\frac{d\Lambda}{dr}+\frac{d\Phi}{dr}
\label{eq:antitrace}
\eeq
%
The first equation is consistent with the Tolman relation.
Eq. (\ref{eq:antitrace}) is the central result of this
paper as articulated by Theorem 3.
One can further verify that Eq.
(\ref{eq:antitrace}) is related to the linear combination of Einstein's
equations given by
\beq
G_{\hat{t}\hat{t}}-8\pi T_{\hat{t}\hat{t}}
+
G_{\hat{r}\hat{r}}-8\pi T_{\hat{r}\hat{r}}
=0 \s .
\eeq
Corollary 1 is proved by verifying that $\Phi$ and $\Lambda$ as
obtained from Einstein's equations (\ref{eq:gtt}) and (\ref{eq:grr})
satisfy Eq. (\ref{eq:antitrace}).  The entropy will therefore
be an extremum.
Corollary 2 can be verified by
substituting the solution from the $G_{\t\t}$ component of Einstein's
equation (Eq. \ref{eq:gtt})
into Eq. (\ref{eq:antitrace})
yielding the correct result for the $G_{\r\r}$ component of
Einstein's equation (Eq. \ref{eq:grr})
%

\section{A thermodynamic sector of quantum gravity \label{therm}}

We now see that for this system, we can recast Einstein's equations
(\ref{eq:gtt})-(\ref{eq:grr})
into another set of equations, namely equation (\ref{eq:antitrace})
which can be classified as thermodynamical in nature, and an additional
equation representing the true physical theory.  The choice of this latter
equation is somewhat arbitrary.
We could for example choose the constraint (Eq. (\ref{eq:constraint}))
to represent the true degrees of freedom.
This equation completely determines $\Lambda$ in terms of the matter
fields.  $\Phi$ is
then determined by extremizing the entropy to find the most probable
configuration. Quantization of the full system is then relatively
straightforward, as the quantization of the gravitational variable $\Lambda$
follows by reducing the phase space via Eq. (\ref{eq:constraint}).  The
full Lagrangian is then given by integrating the Lagrangian density of the
matter fields over the proper volume element
\beq
L=4\pi\int {\cal L}\,r^2 e^{\Phi(r)}\sqrt{1-2 \hat{m}/r}\,dr
\eeq
where $\hat{m}$ is the operator version of $m$ defined via equation
(\ref{eq:m}).  $\Phi$ simply acts as a potential as far as the
quantization of the matter fields go.

The feature that $\Lambda$
is quantum in nature while $\Phi$ is purely thermodynamical
results from the assumption that
equation (\ref{eq:constraint}) gave the physical theory, thus breaking
the symmetry between space and time.

%
We could also choose
for the physical theory, a generally convariant equation
such as the trace of Einstein's equations
\beq
R=\frac{16}{2-d}\pi T  \s.
\eeq
Such an equation and it's quantized version, have appeared at various times
in the literature, particularly in the context of lower dimensional gravity
\cite{trace}.  The physical theory would then give rise to
Eq. (\ref{eq:antitrace}) through Theorem \ref{th1}.

An analogy can be made between our recasting of Einstein's equations into
two sectors, and the thermodynamics of an ideal gas in a box.
Consider for example the derivation of equilibrium conditions for
the gas.
One can derive the condition that the pressure must be constant by imagining
that the box is divided into two sections by a movable, heat conducting wall.
The total volume is fixed, but the volume of each compartment can change as the
wall moves. Extremizing the entropy then gives the conditions for mechanical
equilibrium - namely, that the pressure in each section will be equal. The laws
governing the individual gas molecules and walls of the box can be quantized in
the usual way, but the equation for the equality of pressures should not be
quantized.  The latter equation is just a statement about the most probably
configuration of the system.

Likewise, with our gravitating fluid, one might regard Eq.
(\ref{eq:antitrace}) as coming purely from thermodynamics.  If this is the case,
it should not be quantized.

Although one can divide Einstein's equations into a pure thermodynamic sector
and a physical sector for this particular system,
whether it is
possible to do so more generally as one might conjecture is
speculation.  For a system of thin spherical shells, the
conjecture holds \cite{inprep}, but more work is needed to ascertain how
general these results are.  Certainly
systems which are far out of equilibrium (such as matter containing shock-waves)
would be difficult to analyze, since at present, not enough is known about
non-equilibrium thermodynamics. However, even in the simple model we have
considered, the fact that the entropy is extremized only if the combination of
Einstein's equations given by (\ref{eq:antitrace}) holds is rather puzzling, and
does appear to demand an explanation.

\section{Conclusion}

We have seen that the entropy of a spherically symmetric
fluid in equilibrium is only an extremum when Einstein's equations are
satisfied. Conversely, one can derive part of Einstein's equations
from thermodynamical consideration.  This allows us to recast
Einstein's equations into a smaller physical sector which ought to be
quantized.  This physical sector then gives rise to
equilibrium conditions which are equivalent to the rest of Einstein's equation.
These latter equations ought not to be quantized as they are purely
thermodynamical in nature.  One
can therefore speculate that in general not all of Einstein's equations
should be quantized.

Speculation aside, the identity between Einstein's equations and thermodynamics
for this system is very puzzling, and in many respects analogous to the
connections that exist for black holes. Here however, we have no horizon, and so
the entropy is unambiguously related to the number of states of the system.
In the black-hole case, the entropy is usually associated with the gravitational
degrees of freedom of the horizon.
Furthermore, the connection appears very strong - the condition
for thermodynamic equilibrium
is equivalent to one of Einstein's equations.  This might give some
further insight into the thermodynamics of black holes where the connection
between Einstein's equation and the entropy is more ambiguous.

In fact, one
already sees from equations (\ref{eq:svsr}) and (\ref{eq:gdcurved}) that the
total entropy $S$ need not be purely volume scaling due to
the presence of the metric functions.  The total integrated entropy scales
differently than the total energy as given by equation (\ref{eq:adm}).
This is similar to black holes, where the entropy is also not volume
scaling (being proportional to the area).
This effect was explored in reference
\cite{scaling}, where it was shown that an equilibrium system which has an
extensive entropy in the absence of gravity, will have an entropy proportional
to its area in the limit that the system is about to form a black hole.

The fact that one can calculate the entropy for this system, and that this
entropy is an extremum if Einstein's equations holds, is also of interest in
terms of understanding the thermodynamics of strongly interacting systems.
It appears that one can learn much about gravitational systems simply
by conducting the analysis in terms of local thermodynamical
variables.
%
It is hoped that the results here might
lead to a greater understanding of other non-extensive systems.

%
%
%
%
%

\vskip 2cm \noindent{\bf
Acknowledgments:} It is a pleasure to thank Don Page for many
interesting and valuable discussions.  It is also a pleasure to thank Robb Mann,
Sumati Surya, Ivan Booth, Eric Poisson, and those at P.I. for their helpful
comments.  This research was supported in part by NSERC Canada.

\end{document}